\documentclass[preprint,showpacs,preprintnumbers,amsmath,amssymb]{revtex4}

\usepackage{epsfig}

\usepackage{amssymb}
\usepackage{amsmath}

\begin{document}

\title{Universal amplitude ratios for scaling corrections on Ising strips
with fixed boundary conditions.}

\author{N. Sh. Izmailian}
\email{izmailan@phys.sinica.edu.tw}
\affiliation{Institute of Physics, Academia Sinica, Nankang, Taipei 11529, Taiwan}
\affiliation{Yerevan Physics Institute, Alikhanian Brothers 2, 375036 Yerevan, Armenia}

\date{\today}

\begin{abstract}
We study the (analytic) finite-size corrections in the Ising model
on the strip with fixed ($+ -$) boundary conditions. We find that
subdominant finite-size corrections to scaling should be to the
form $a_k/N^{2k-1}$ for the free energy $f_N$ and
$b_k^{(n)}/N^{2k-1}$ for inverse correlation length $\xi_n^{-1}$,
with integer value of $k$. We investigate the set $\{a_k,
b_k^{(n)}\}$ by exact evaluation and their changes upon varying
anisotropy of coupling. We find that the amplitude ratios
$b_k^{(n)}/a_k$ remain constant upon varying coupling anisotropy.
Such universal behavior are correctly reproduced by the conformal
perturbative approach.
\end{abstract}

\maketitle

\section{Introduction}
\label{introduction}

Finite-size scaling and corrections for critical systems have
attracted much attention in recent decades. Although many
theoretical results are now known about the critical exponents and
universal relations among the leading critical amplitudes, not
much information is available on ratios among the amplitudes in
finite-size correction terms \cite{fishercorr}. New universal
amplitude ratios have been recently presented for the Ising model
\cite{izmailian2001,izmailian2009a,izmailian2009b,izmailian2010}.
Consider an Ising ferromagnet on an $N \times M$ lattice. If
$\Lambda_0 > \Lambda_1 > \Lambda_2 > \Lambda_3 > ... $ are the
eigenvalues of the transfer matrix (TM), in the limit $M \to
\infty$ the free energy per spin, $f_N$, and the inverse
longitudinal spin-spin correlation length, $\xi_n^{-1}$, are
\begin{equation}
f_N=\frac{1}{\zeta N} \ln{\Lambda_0} \quad \mbox{and} \quad
\xi_n^{-1}=\frac{1}{\zeta}\ln{(\Lambda_0/\Lambda_n)}. \label{I}
\end{equation}
Here $\zeta$ is a geometric factor, which is unity for the square
lattice and, in triangular and honeycomb geometries (also for the
square lattice when the TM progresses along the diagonal
\cite{OBrien}), corrects for the fact that the physical length
added upon each application of the TM differs from one lattice
spacing \cite{Privman}.

At the critical point $T_c$ the asymptotic finite-size scaling
behavior of the critical free energy ($f_N$) and the inverse
correlation lengths ($\xi_n^{-1}$) of an infinitely long 2D strip
of finite width $N$ has the form \cite{Blote,Blote1}
\begin{equation}
\lim_{N \to \infty} {N^2 (f_N-f_{\infty})- 2 N f_{surf}}=A,
\label{I2}
\end{equation}
\begin{equation}
\lim_{N \to \infty} {N \xi_n^{-1}}=D_n, \label{I1}
\end{equation}
where $f_{\infty}$ is the bulk free energy, $f_{surf}$ is the
surface free energy and $A$ and $D_n$ are the universal constants,
but may depend on the boundary conditions. In some 2D geometries,
the values of $A$ and $D_n$ are known
\cite{Blote,Blote1,cardy86a}, to be related to the conformal
anomaly number ($c$), the conformal weight of the ground state
$(\Delta)$, and  the scaling dimension of the $n$-th scaling field
($x_n$) of the theory
\begin{eqnarray}
A&=& \pi \zeta \left(\frac{c}{24}-\Delta\right),  \qquad \quad D_n
= \pi \zeta x_n, \label{Afree}
\end{eqnarray}
for strip geometry. Here $\zeta $ is anisotropy parameter. The
principle of unitarity of the underlying field theory restricts
through the Kac formula the possible values of c and for each
value of c only permits a finite number of possible values of
$\Delta$. For the 2D Ising model, we have $c=1/2$ and the only
possible values are $\Delta= 0, 1/16, 1/2$.

For Ising universality class there are five different boundary
universality classes: periodic, antiperiodic, free, fixed and
mixed (mixture of the last two). Two of them: periodic and
antiperiodic boundary universal classes are in cylinder geometry
and three boundary universal classes: free, fixed ($+ -$) and
mixed in strip geometry. For fixed ($+ -$) boundary conditions the
spins are fixed to the opposite values on two sides of the strip.
The mixed boundary conditions corresponds to free boundary
conditions on one side of the strip, and fixed boundary conditions
on the other. In the terminology of surface critical phenomena
these three boundary universal classes: free, fixed ($+-$) and
mixed correspond to "ordinary", "extraordinary" and "special"
surface critical behavior, respectively.

The highest conformal weight $\Delta$, and the scaling dimension
$x_n$ depends on the boundary conditions and for fixed $(+-)$
boundary conditions they given by
\begin{eqnarray}
\Delta&=& \frac{1}{2}, \hspace{1cm} x_1=1 ,\quad x_2=2, \quad
x_3=3, ..., \quad x_n=n. \label{fixedD}
\end{eqnarray}

Quite recently, Izmailian and Hu \cite{izmailian2001} studied the
finite size correction terms for the free energy per spin and the
inverse correlation lengths of critical 2D Ising models  on $N
\times \infty$ lattice and 1D quantum Ising chain with periodic
boundary conditions. They obtained analytic expressions for the
finite-size correction coefficients $a_k$ and $b_k^{(n)}$ in the
expansions
\begin{eqnarray}
N [f_N-f_{\infty}] &=&f_s+\sum_{k=1}^{\infty}\frac{a_k}{N^{2
k-1}},
\label{fN} \\
\xi_n^{-1}&=&\sum_{k=1}^{\infty}\frac{b_k^{(n)}}{N^{2 k-1}},
\label{cli}
\end{eqnarray}
and find that although the finite-size correction coefficients
$a_k$, $b_k^{(n)}$ (for $n=1,2$) themselves are non-universal
(except for $a_1$ and $b_1^{(n)}$), the amplitude ratios for the
coefficients of these series $b_k^{(n)}/a_k$  are universal.

Later that result has been extended  for the quantum Ising chain
for antiperiodic and free boundary conditions
\cite{izmailian2009a} and for the two-dimensional (2D) Ising
models on ${\cal M} \times \infty$ lattice with the special
boundary conditions studied by Brascamp and Kunz (BK)
\cite{izmailian2009b}. It was shown that Brascamp and Kunz (BK)
 boundary conditions \cite{Brascamp} belong to the mixed
boundary condition universality class although the mixed boundary
condition and the BK boundary condition are different on one side
of the long strip.

In the present paper we will present exact calculations for a set
of universal amplitude ratios for the Ising model with fixed
$(+-)$ boundary conditions. We obtain analytic equations for $a_k$
and $b_k^{(n)}$ (for $n=1,2,3,...$) in the expansions given by
Eqs. (\ref{fN}) and (\ref{cli}) and find that amplitude ratios
$b_k^{(n)}/a_k$ are universal. We will show that such universal
behavior are correctly reproduced by the conformal perturbative
approach.

\section{The Ising model on cylinder}
\label{ising} The Ising model on a cylinder, with width N and
height M and free, fixed and mixed boundary condition along
diagonal was formulated in \cite{OBrien} in terms of a double-row
transfer matrix D(u). They define the lattice ${\cal L}$ as a
square lattice rotated by 45 degrees, in which the rows have
alternately $L-1$ and $L$ faces.  Vertically the lattice have
columns of $L'$ faces, and the periodic boundary conditions
imposed in this direction by identifying the first and $(L'+1)$th
rows of faces. The lattice ${\cal L}$ consists of $N$ (zigzagging)
columns edges and $M$ (zigzagging) rows edges, respectively, and
$N$ and $M$ are given by
\begin{eqnarray}
N=2L \qquad M=2L' \label{LMN}
\end{eqnarray}
The lattice ${\cal L}$ is shown in Fig. 1. We are concerned with
the fixed $(+-)$ type of boundary conditions. For fixed $(+-)$
boundary conditions we choose the spins at the left and right
boundaries of the lattice ${\cal L}$ to be $+ 1$ and $- 1$.

\begin{figure}
\epsfxsize=90mm \vbox to3in{\rule{0pt}{3in}}
\includegraphics{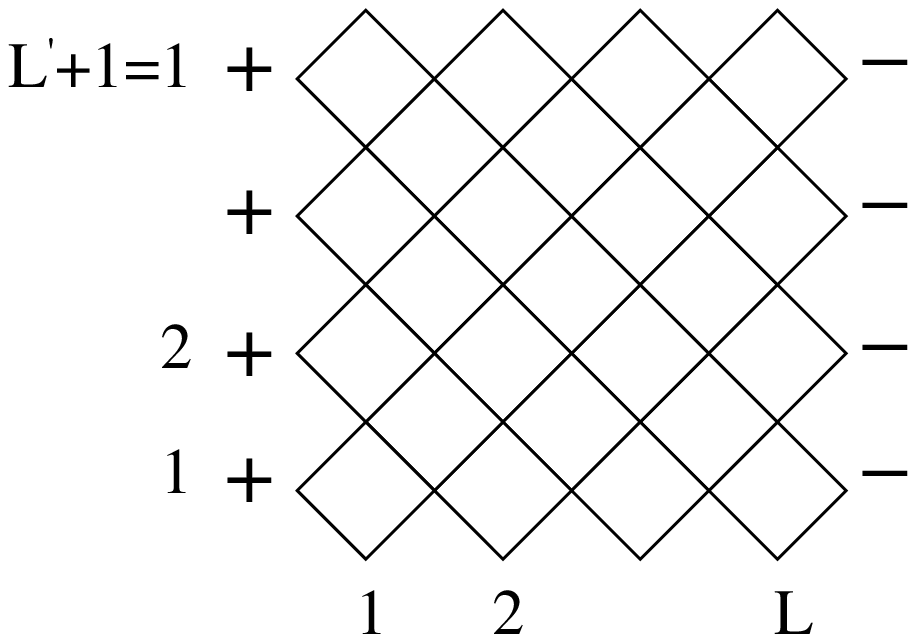}  \caption{The lattice ${\cal L}$}
\end{figure}

The partition function is given by
\begin{equation}
Z_{NM}=\sum_{\{\sigma\}}\exp{\left(J\sum_{<i,j>}\sigma_i
\sigma_j+K\sum_{<i,j>}\sigma_i \sigma_j\right)}
\end{equation}
where we define the set of spins not fixed by the boundary
conditions by $\{\sigma\}$. For fixed  boundary conditions, the
first sum is over edges in odd columns and the second sum over
edges in even columns of the lattice.

Consider the critical Ising model on a cylindrical lattice of $N$
columns and $M$ rows as described previously. The finite-size
partition function $Z_{N M}$ can be written as
\begin{equation}
Z_{N M}(u)=Tr\left(D(u)^M\right)=\sum_n e^{-2M E_n(u)}\label{ZNM}
\end{equation}
where the sum is over all eigenvalues of a transfer matrix $D(u)$,
written as $e^{-2 E_n(u)}$. Since we restrict ourselves to the
critical Ising model, we have $\sinh(2J) \sinh( 2K) = 1$. This
condition can be conveniently parameterized by introducing a
so-called spectral parameter $u$, so that $\sinh(2J) = \cot(2u),
\sinh(2K) = \tan(2u)$, with $0 < u < \pi/4$. The anisotropy
parameter $\zeta$ related to the spectral parameter $u$ through
\begin{equation}
\zeta=\sin{4u} \label{zeta}
\end{equation}
For isotropic system ($K=J$) we have $u=\pi/8$ and $\zeta=1$.

For fixed $(+-)$ boundary conditions all eigenvalues have been
determined in \cite{OBrien}, for any finite value of N. The energy
$E_n$ associated with the eigenvalues of a transfer matrix $D(u)$
are given by
\begin{eqnarray}
E_n(u)=-\frac{1}{2}\ln\left\{2^{-N/2}\prod_{k=1}^{N/2}\left[{\rm
cosec} \left(\frac{k-1/2}{N+1}\pi\right)+\mu_k
\sin(4u)\right]\right\} \label{lambda}
\end{eqnarray}
Using the identity
\begin{equation}
\prod_{k=1}^{L}\sin\left(\frac{k-1/2}{2L+1}\pi\right)=2^{-L}
\end{equation}
the Eq. (\ref{lambda}) can be simplified to the following form
\begin{eqnarray}
E_n(u)=-\frac{1}{2}\sum_{k=1}^{N/2}\ln\left[1+\mu_k \sin(4u)\sin
\left(\frac{k-1/2}{N+1}\pi\right)\right] \label{lambda1}
\end{eqnarray}
where $\mu_k=\pm 1$.  For fixed  $(+-)$ boundary conditions
$\mu_k$ should satisfied the conditions $\prod_{k=1}^L\mu_k=-1$.
This correctly yields $2^{L-1}$ eigenvalues for  the fixed $(+-)$
boundary cases.

The ground state $E_0$ correspond to all $\mu_k=1$ except
$\mu_1=-1$ in  Eq. (\ref{lambda1}). The exited levels $E_n$ are
given by  Eq. (\ref{lambda1}) with $\mu_n=-1$ and all other
$\mu_k$ are equal to 1: $\mu_k=1$ ($k \ne n$). Thus for the
critical free energy ($N f_N=-E_0$) and the inverse correlation
lengths ($\xi_r^{-1}=E_r-E_0$) (for $r=1,2,...,n $) we obtain
\begin{eqnarray}
N f_N&=&\sum_{k=1}^{\frac{N}{2}}\omega
\left(\frac{k-1/2}{N+1}\pi\right)+\omega
\left(-\frac{\pi}{2(N+1)}\right)-\omega
\left(\frac{\pi}{2(N+1)}\right),\label{freefN}\\
\xi_1^{-1}&=&\omega \left(\frac{3\pi}{2(N+1)}\right)-\omega
\left(\frac{\pi}{2(N+1)}\right)+\omega
\left(-\frac{\pi}{2(N+1)}\right)-\omega
\left(-\frac{3\pi}{2(N+1)}\right),\label{xif1}\\
\xi_2^{-1}&=&\omega \left(\frac{5\pi}{2(N+1)}\right)-\omega
\left(\frac{\pi}{2(N+1)}\right)+\omega
\left(-\frac{\pi}{2(N+1)}\right)-\omega
\left(-\frac{5\pi}{2(N+1)}\right),\label{xif2}\\
\xi_3^{-1}&=&\omega \left(\frac{7\pi}{2(N+1)}\right)-\omega
\left(\frac{\pi}{2(N+1)}\right)+\omega
\left(-\frac{\pi}{2(N+1)}\right)-\omega
\left(-\frac{7\pi}{2(N+1)}\right),\label{xif3}\\
&\vdots& \nonumber\\
\xi_n^{-1}&=&\omega \left(\frac{(2n+1)\pi}{2(N+1)}\right)-\omega
\left(\frac{\pi}{2(N+1)}\right)+\omega
\left(-\frac{\pi}{2(N+1)}\right)-\omega
\left(-\frac{(2n+1)\pi}{2(N+1)}\right),\label{xifn}
\end{eqnarray}
where $\omega(x)$ is given by
\begin{equation}
\omega(x)=\frac{1}{2}\ln{\left[1+\sin(4u)\sin x\right]}
\end{equation}
The sum in Eq. (\ref{freefN}) can be transformed as
\begin{eqnarray}
\sum_{k=1}^{\frac{N}{2}}\omega
\left(\frac{k-1/2}{N+1}\pi\right)=-\frac{1}{2}\omega
\left(\frac{\pi}{2}\right)+\frac{1}{2}\sum_{k=0}^{N}\omega
\left(\frac{k+1/2}{N+1}\pi\right),\label{freefN1}
\end{eqnarray}
and can be handled by using the Euler-Maclaurin summation formula
\cite{hardy}. Suppose that $F(x)$ together with its derivatives is
continuous within the interval $(a, b)$. Then the general
Euler-Maclaurin summation formula states
\begin{equation}
\sum_{n=0}^{N-1} F(a+n h+\alpha h)=\frac{1}{h}\int_{a}^{b}
F(\tau)~{\rm d}\tau + \sum_{k=1}^{\infty} \frac{h^{k-1}}{k!} {\rm
B}_{k}(\alpha)\left(F^{(k-1)}(b)-F^{(k-1)}(a)\right)
\label{EMFormula}
\end{equation}
where $0 \le \alpha \le 1$, $h=(b-a)/N$ and ${\rm B}_{k}(\alpha)$
are so-called Bernoulli polynomials defined in terms of the
Bernoulli numbers $B_p$ by
\begin{equation}
{\rm B}_{k}(\alpha)=\sum_{p=0}^{k}{\rm B}_{p}\frac{k!}{(k-p)!p!}
\alpha^{k-p}
\end{equation}
Indeed, $B_n(0)=B_n$. Bernoulli polynomials satisfy the identity:
\begin{equation}
B_n(1/2)=\left(2^{1-n}-1\right)B_n \label{bernoulli}
\end{equation}
Using the Euler-Maclaurin summation formula given by Eq.
(\ref{EMFormula}) with $a = 0, b = \pi, \alpha = 1/2$ and
 $F(x)=\omega(x)$ the asymptotic expansion of the critical free
energy $f_N$ can be written in the following form
\begin{eqnarray}
N(f_N - f_{\infty}) &=&f_{surf}-
\sum_{p=1}^{\infty}\frac{\lambda_{2p-1}}{(2p-1)!}\left(\frac{B_{2
p}(1/2)}{2p}+2^{-2p+2}\right)
\left(\frac{\pi}{N+1}\right)^{2 p-1}, \label{fNfixed2}\\
&=&f_{surf}-\frac{23\pi \sin {4u}}{48(N+1)}
-\frac{247(\sin^3{4u}-\frac{1}{2}\sin{4u})}{5760}
\left(\frac{\pi}{N+1}\right)^3 +\dots, \nonumber
\end{eqnarray}
where
\begin{eqnarray}
f_{\infty}&=&\frac{1}{\pi}\int_0^{\pi}\omega(x)dx
\label{fb}\\
f_{surf}&=&-\frac{1}{4}\ln{(1+\sin{4u})}
+\frac{1}{\pi}\int_0^{\pi}\omega(x)dx\label{f1}
\end{eqnarray}
and $\lambda_{k}$ is the coefficients in
 the Taylor expansion of the $\omega(x)$:
\begin{eqnarray}
\omega(x)&=&\sum_{p=0}^{\infty} \frac{\lambda_{p}}{p!}\;x^{p},
\label{SpectralFunctionExpansion1}
\end{eqnarray}
with $\lambda_0=0, \lambda_1=\frac{1}{2}\sin{4u},
\lambda_2=-\frac{1}{2}\sin{4u}^2,
\lambda_3=\sin^3{4u}-\frac{1}{2}\sin{4u}, ... $.

Using the Taylor expansion of the $\omega(x)$ the asymptotic
expansion of the critical inverse correlation lengths $\xi_r^{-1}$
for $r=1, 2, 3, ..., n$ can be written  in the following form
\begin{eqnarray}
\xi_1^{-1}&=&\sum_{p=1}^{\infty}\frac{2(3^{2p-1}-1)\lambda_{2p-1}}{2^{2p-1}(2p-1)!}
\left(\frac{\pi}{N+1}\right)^{2p-1}
, \label{xi1f2}\\
&=&\frac{\pi \sin {4u}}{N+1}
+\frac{13(\sin^3{4u}-\frac{1}{2}\sin{4u})}{12}
\left(\frac{\pi}{N+1}\right)^3 +\dots, \nonumber\\
\xi_2^{-1}&=&\sum_{p=1}^{\infty}\frac{2(5^{2p-1}-1)\lambda_{2p-1}}{2^{2p-1}(2p-1)!}
\left(\frac{\pi}{N+1}\right)^{2p-1}
, \label{xi2f2}\\
&=&\frac{2\pi \sin {4u}}{N+1}
+\frac{31(\sin^3{4u}-\frac{1}{2}\sin{4u})}{6}
\left(\frac{\pi}{N+1}\right)^3 +\dots, \nonumber\\
\xi_3^{-1}&=&\sum_{p=1}^{\infty}\frac{2(7^{2p-1}-1)\lambda_{2p-1}}{2^{2p-1}(2p-1)!}
\left(\frac{\pi}{N+1}\right)^{2p-1}
, \label{xi3f2}\\
&=&\frac{3\pi \sin {4u}}{N+1}
+\frac{57(\sin^3{4u}-\frac{1}{2}\sin{4u})}{4}
\left(\frac{\pi}{N+1}\right)^3 +\dots, \nonumber\\
&\vdots& \nonumber\\
\xi_n^{-1}&=&\sum_{p=1}^{\infty}\frac{2[(2n+1)^{2p-1}-1]\lambda_{2p-1}}{2^{2p-1}(2p-1)!}
\left(\frac{\pi}{N+1}\right)^{2p-1}
, \label{xinf2}\\
&=&\frac{n \pi \sin {4u}}{N+1}
+\frac{[(2n+1)^3-1](\sin^3{4u}-\frac{1}{2}\sin{4u})}{24}
\left(\frac{\pi}{N+1}\right)^3 +\dots, \nonumber
\end{eqnarray}
Eqs. (\ref{fNfixed2}) and (\ref{xi1f2}) - (\ref{xinf2}) imply that
the ratios of amplitudes of the $(N+1)^{-(2p-1)}$ correction terms
in the free energy and the inverse correlation lengths expansions,
i.e., $b_p^{(n)}/a_p$ should not depend on the spectral parameter
$u$ and given by
\begin{eqnarray}
\frac{b_p^{(n)}}{a_p}&=&-\frac{(2n+1)^{2p-1}-1}{\frac{2^{2p-2}B_{2p}(1/2)}{2p}+1}.
\label{ratio}
\end{eqnarray}
For $p=1$ we have
\begin{eqnarray}
\frac{b_1^{(n)}}{a_1}&=&-\frac{48}{23}n \label{ratio1}
\end{eqnarray}
and for $p=2$ we have
\begin{eqnarray}
\frac{b_2^{(n)}}{a_2}&=&-\frac{240}{247}\left[(2n+1)^3-1\right]
=-\frac{480}{247}n(4n^2+6n+3). \label{ratio2}
\end{eqnarray}
\section{Perturbated conformal field theory}
The finite-size corrections to Eqs. (\ref{I2}) and (\ref{I1}) can
be calculated by the means of a perturbated conformal field theory
\cite{cardy86,zamol87}. In general, any lattice Hamiltonian will
contain correction terms to the critical Hamiltonian $H_c$
\begin{equation}
H = H_c + \sum_p g_p \int_{-N/2}^{N/2}\phi_p(v) d v, \label{Hc}
\end{equation}
where $g_p$ is a non-universal constant and $\phi_p(v)$ is a
perturbative conformal field. Below we will consider the case with
only one perturbative conformal field, say $\phi_l(v)$. Then the
eigenvalues of $H$ are
\begin{equation}
E_n=E_{n,c}+  g_l \int_{-N/2}^{N/2}<n|\phi_l(v)|n> d v + \dots,
\label{En}
\end{equation}
where $E_{n,c}$ are the critical eigenvalues of $H$. The matrix
element $<n|\phi_l(v)|n>$ can be computed in terms of the
universal structure constants $(C_{nln})$ of the operator product
expansion \cite{cardy86}: $<n|\phi_l(v)|n> =\left({2
\pi}/{N}\right)^{x_l}C_{nln}$, where $x_l$ is the scaling
dimension of the conformal field $\phi_l(v)$. The energy gaps
$(E_n-E_0)$ and the ground-state energy ($E_0$) can be written as
\begin{eqnarray}
E_n-E_0&=&\frac{2 \pi}{N} x_n+ 2 \pi
g_l(C_{nln}-C_{0l0})\left(\frac{2 \pi}{N}\right)^{x_l-1} + \dots,
\label{xin}\\
E_0 &=& E_{0,c}+2 \pi  g_l C_{0l0} \left(\frac{2
\pi}{N}\right)^{x_l-1} + \dots. \label{E0conf}
\end{eqnarray}
Note, that the ground state energy $E_0$ and the energy gaps
($E_n-E_0$) of a quantum spin chain are, respectively, the quantum
analogies of the free energy $f_N$ and inverse correlation lengths
$\xi_n^{-1}$ for the Ising model; that is,
\begin{equation}
N f_{N} \Leftrightarrow - E_0,  \quad \mbox{and} \quad \xi_n^{-1}
\Leftrightarrow E_n-E_0 . \label{definition}
\end{equation}
For the 2D Ising model, one finds \cite{henkel} that the leading
finite-size corrections ($1/N^3$) can be described by the
Hamiltonian given by Eq. (\ref{Hc}) with a single perturbative
conformal field $\phi_l(v)=L_{-2}^2(v)$ with scaling dimension
$x_l=4$ .

In order to obtain the corrections we need the matrix elements
$<n|L_{-2}^2(v)|n>$, which have already been computed by Reinicke
\cite{reinicke87}.
\begin{eqnarray}
<\Delta+r|L_{-2}^2|\Delta+r> &=&
\left(\frac{2\pi}{N}\right)^4\left[\frac{49}{11520}+(\Delta+r)\left(\Delta-\frac{5}{24}
+\frac{r(2 \Delta + r)(5
\Delta+1)}{(\Delta+1)(2\Delta+1)}\right)\right]
\label{matrixelement}
\end{eqnarray}
The universal structure constants $C_{2l2}$, $C_{1l1}$ and
$C_{0l0}$ can be obtained from the matrix element
$<n|L_{-2}^2(v)|n> =\left({2 \pi}/{N}\right)^{x_l}C_{nln}$, where
$x_l = 4$ is the scaling dimension of the conformal field
$L_{-2}^2(v)$.

At the critical point  the spectra of the Hamiltonian with  fixed
$(+-)$ boundary conditions  can be understood in terms of
irreducible representations $\Delta$ of a single Virasoro algebra
with values of $\Delta$ is $\frac{1}{2}$.

For fixed ($+ -$) boundary conditions the ground state $|0>$ and
the excited states $|r>$ ($r=1, 2, ..., n$) are given by
\begin{eqnarray}
|r>&=&|\Delta=\frac{1}{2}, r=k>, \qquad \mbox{for} \qquad  k = 1,
2, ..., n \label{state0f1}
\end{eqnarray}
After reaching this point, one can easily compute the universal
structure constants $C_{rlr}$ ($r=0,1,2,...,n$) for fixed $(+-)$
boundary conditions. The values of $C_{rlr}$ can be obtained from
Eqs. (\ref{matrixelement}) and (\ref{state0f1}) and given by:
\begin{eqnarray}
C_{rlr}&=&\frac{1729}{11520}+\frac{7r(4r^2+6r+3)}{24} \nonumber\\
C_{0l0}&=&1729/11520, \qquad C_{1l1} = 45409/11520, \qquad
C_{2l2}=210049/11520, .... \label{Cm1}
\end{eqnarray}

Equations (\ref{xin}) and (\ref{E0conf}) implies that the ratios
of first-order corrections amplitudes for $E_n-E_0$ ($\xi_n^{-1}$)
and $- E_0$ ($f_{N}$) is universal and equal to
$(C_{0l0}-C_{nln})/C_{0l0}$, which is consistent with Eq.
(\ref{ratio2})
\begin{eqnarray}
r_n(2)&=&\frac{C_{0l0}-C_{nln}}{C_{0l0}}
=-\frac{480}{247}n(4n^2+6n+3)\label{rnfixed}
\end{eqnarray}

\section{Summary}

In this paper we present exact calculations for all coefficients
in the asymptotic expansions given by Eqs. (\ref{fN}) and
(\ref{cli}) for an infinite number of specific levels $E_n$
($n=0,1,2,...$), for which we find that the ratios $b_p^{(n)}/a_p$
are indeed universal and doesn't depend on the spectral parameter
$u$. We find that such universal behavior are correctly reproduced
by the conformal perturbative approach.

\section{Acknowledgements}

The work of NSI is supported in part by National Center for
Theoretical Sciences: Physics Division, National Taiwan
University, Taipei, Taiwan and by Institute of Physics, Academia
Sinica, Taipei, Taiwan.

\end{document}